\documentclass[useAMS,usenatbib]{mn2e}
 \pdfoutput=1
\usepackage[version=3]{mhchem}

\usepackage{natbib}
\usepackage{graphicx}
\usepackage{lastpage}

\title[Thermal H/D exchange]
  {Thermal H/D exchange in polar ice - deuteron scrambling in space}
\author[T.~Lamberts et al.]
  {T.~Lamberts,$^{1,2}$\thanks{corresponding author} S.~Ioppolo,$^{2}$ H.~M.~Cuppen,$^2$ G~.Fedoseev,$^1$ and H.~Linnartz$^1$\\  
  $^1$~Raymond and Beverly Sackler Laboratory for Astrophysics, Leiden Observatory, Leiden University, \\\hspace{10pt}P.O. Box 9513, NL 2300 RA Leiden, The Netherlands.\\
  $^2$~Faculty of Science, Radboud University Nijmegen, IMM, \\\hspace{10pt}P.O. Box 9010, NL 6500 GL Nijmegen, The Netherlands}
\date{Released 2014 Xxxxx XX}

\pagerange{\pageref{firstpage}--\pageref{lastpage}} \pubyear{2014}

\def\LaTeX{L\kern-.36em\raise.3ex\hbox{a}\kern-.15em 
    T\kern-.1667em\lower.7ex\hbox{E}\kern-.125emX}

\begin{document}

\label{firstpage}

\maketitle

\begin{abstract}
We have investigated the thermally induced proton/deuteron exchange in mixed amorphous \ce{H2O}:\ce{D2O} ices by monitoring the change in intensity of characteristic vibrational bending modes of \ce{H2O}, \ce{HDO}, and \ce{D2O} with time and as function of temperature. The experiments have been performed using an ultra-high vacuum setup equipped with an infrared spectrometer that is used to investigate the spectral evolution of homogeneously mixed ice upon co-deposition in thin films, for temperatures in the 90 to 140~K domain. With this non-energetic detection method we find a significantly lower activation energy for H/D exchange -- $3840 \pm 125$~K -- than previously reported. Very likely this is due to the amorphous nature of the interstellar ice analogues involved. This provides reactive timescales ($\tau<10^4$ years at $T$ $>70$~K) fast enough for the process to be important in interstellar environments. Consequently, an astronomical detection of \ce{D2O} will be even more challenging because of its potential to react with \ce{H2O} to form HDO. Furthermore, additional experiments, along with previous studies, show that proton/deuteron swapping also occurs in ice mixtures of water with other hydrogen bonded molecules, in particular on the OH and NH moieties. 
We conclude that H/D exchange in ices is a more general process that should be incorporated into ice models that are applied to protoplanetary disks or to simulate the warming up of cometary ices in their passage of the perihelion, to examine the extent of its influence on the final deuteron over hydrogen ratio. 
\end{abstract}


\section{Introduction}
The delivery of water to Earth is far from understood. Several hypotheses have been put forward, among which delivery by comets and/or asteroids during the so-called `late veneer' stage \citep{Morbidelli:2000,Cleeves:2014}, direct adsorption of water onto grains prior to planetary accretion \citep{Muralidharan:2008}, and in the recent `Grand Tack' model, water delivery during the formation phase of terrestrial planets \citep{OBrien:2014}. Since the ratio between deuterons and protons incorporated in water molecules is elevated in the Earth Ocean with respect to that of the initial bulk solar composition, the D/H fraction in various molecules is a logical tracer for the proposed origins. This concerns particularly the ratio of HDO over \ce{H2O}. The \ce{D2O} abundance is expected to be low, because the amount of deuterium present in the interstellar medium is a factor 100 000 less than that of hydrogen. In fact the number of astronomical \ce{D2O} detections is very limited: in the cold envelope layer surrounding solar-type protostar IRAS 16293-2422 \citep{Butner:2007, Vastel:2010, Coutens:2013} and in the warmer regions surrounding the Class 0 protostar NGC 1333 IRAS2A \citep{Coutens:2014}. 

\begin{figure*}
\begin{center}
\includegraphics[width=32mm]{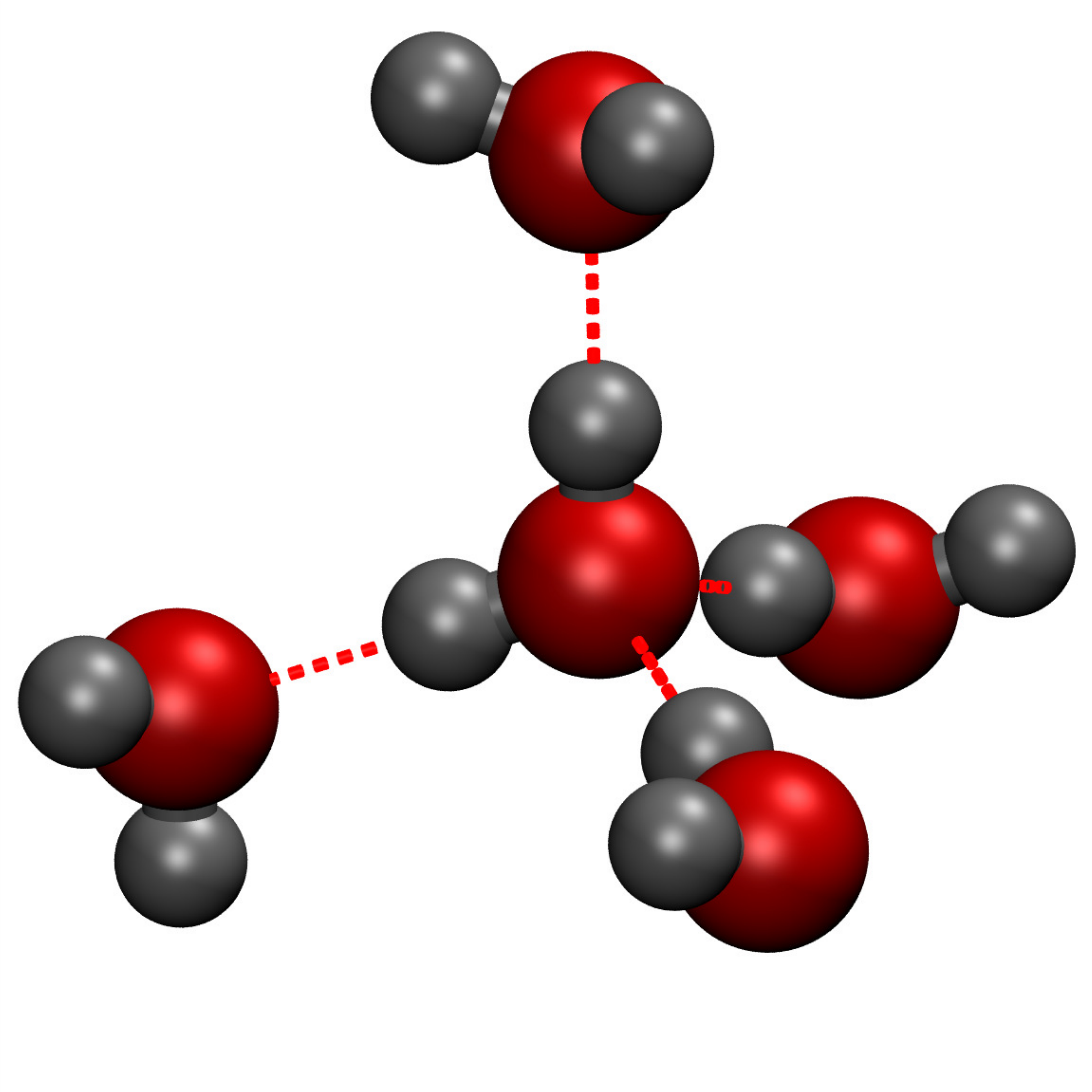}\hspace{20pt}
\includegraphics[width=32mm]{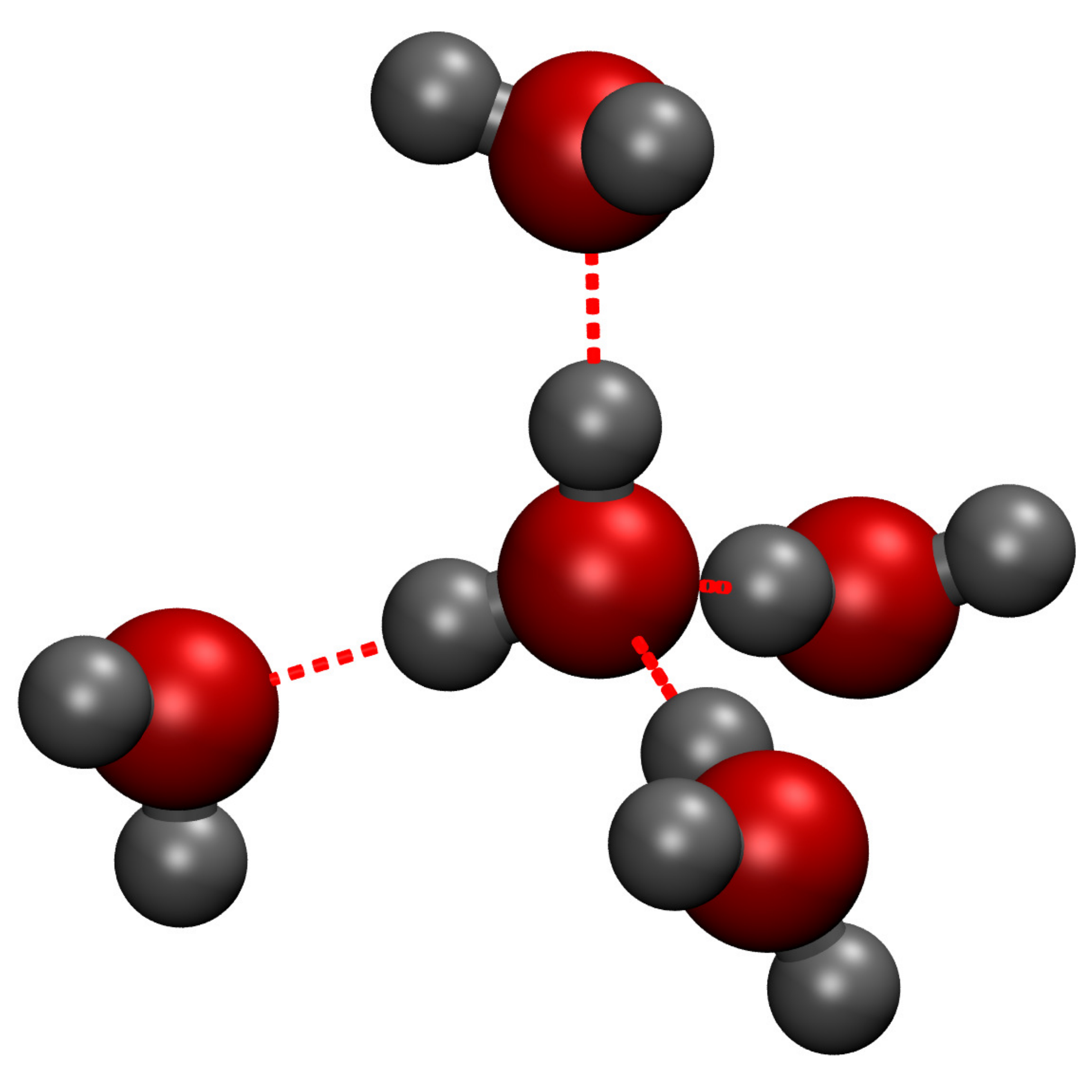}\hspace{20pt}
\includegraphics[width=32mm]{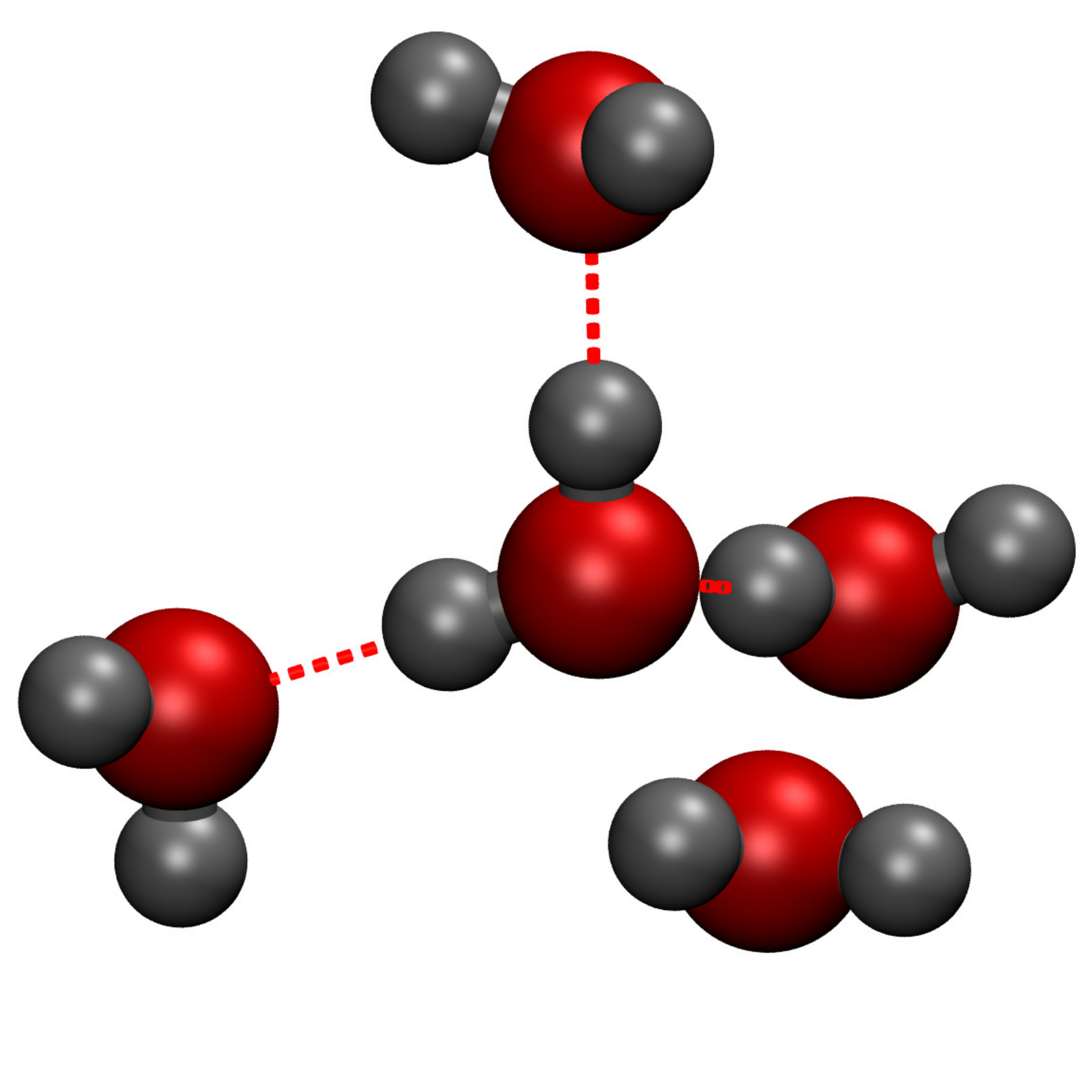}\\
\large a \hspace{105pt} b \hspace{105pt} c
\end{center}
\caption{(a) Water molecule surrounded by four hydrogen bonded molecules as in an ideal crystal structure, (b) water molecule surrounded by four hydrogen bonded water molecules, where one of those contains an extra proton: ionic defect, and (c) water molecule surrounded by four water molecules, with one of those rotated causing a missing hydrogen bond: `L-defect'.}
\label{waterstruct}
\end{figure*}

It should be noted, though, that a direct comparison of observed D/H ratios between different locations may be a too naive approach, to identify the origin of our water, as obviously a number of processes are at play that may scale differently over time for different conditions. For example, investigating molecules other than water, it has been found that deuterium fractionation in low-mass protostellar envelopes and molecular outflows for \ce{H2CO} and \ce{CH3OH} are higher than those for \ce{H2O} and \ce{NH3} as summarized in Fig.~11 by \citet{Caselli:2012}. This difference is argued to reflect the temporal sequence in which the species are formed, as with decreasing gas-phase CO (increasing frozen out solid CO) abundances the gas-phase atomic D/H ratio is found to increase. Final values may also be different in different environments, \emph{e.g.}, low- vs. high-mass protostellar envelopes. In order to interpret molecular D/H ratios correctly, it is important to investigate all impacting processes. In the past, the focus has been on isotope dependent gas-phase chemical processes. So far, the two main routes are first gas-phase isotope dependent reactions with \ce{H2D+} transferring its deuterium atom and second the enhanced gas-phase D/H ratio after CO freeze-out that impacts on ice chemistry by the higher probability of deuterium deposition \citep{Tielens:1983, Caselli:2012}. In the solid state, on the grains, deuterium fractionation is further determined by the chemical surface reaction networks. These involve both hydrogenation and deuteration pathways and the crosslinks between them. This paper focuses in detail on one particular process only: thermally induced proton/deuteron exchange reactions in the ices.

In general, the grain temperature during the formation of the icy mantles in the dense and dark regions of the interstellar medium is low, between 10-20~K \citep{Bergin:2007}. The exact temperature determines the surface reaction routes taken, by, \emph{e.g.}, sticking coefficients and the relative importance of tunneling transmission coefficients, that lead eventually to the formation of species like \ce{CH3OH} and \ce{H2O} \citep{Hiraoka:1998, Watanabe:2002, Fuchs:2009, Miyauchi:2008, Ioppolo:2008}. At later evolutionary stages in the star forming sequence, surface temperatures are no longer homogeneously distributed throughout the core or disk and can range up to 100~K \citep{Tak:2000, Jorgensen:2002, Aikawa:2002, Nomura:2005, Bergin:2007, Launhardt:2013}. At 100~K, ices are expected to have been fully evaporated, thus transferring molecules from the grains to the gas phase. 

Within the astrochemical community, proton exchange in polar (\emph{i.e.}, water-rich) ices has attracted special attention. In the context of ion formation at low and higher temperatures, \citet{Grim:1989} discuss the proton hopping between the hydrogen bonded molecules leading to the formation of \ce{NH4+} and \ce{OCN-}. As stated by \citet{Tielens:2013} heating and sublimation seem to be important in interstellar ices near young stars in modeling the warm gas; non-energetic thermal reactions may therefore be relevant as well. Proton exchange can be important in particular since the main component of ices is \ce{H2O} and, moreover, the main reservoir of deuterium is HDO \citep{Rodgers:2002}. If indeed such a scrambling of protons and deuterons occurs efficiently, this implies that not the low-temperature reaction routes are decisive for final HDO/\ce{H2O} ratio, but rather the total number of H and D atoms incorporated in hydrogen bonded molecules.  \citet{Ratajczak:2009} showed that the hydroxyl group in a methanol molecule exchanges its proton or deuteron with surrounding water molecules and proposes this as one of the possible explanations for the deviation between modeled and observed \ce{CH2DOH}/\ce{CH3OD} ratios \citep{Ratajczak:2011}.

Interstellar or circumstellar ices and cometary ices are thought to have a common chemical origin \citep{Bockelee:2000}, but the temperature processing of ices in comets is of an intrinsically different nature. Ices are typically much thicker, \emph{i.e.}, orders of magnitude compared to the layer thicknesses of several tens of monolayers typical for interstellar ices. Moreover, each passage through the perihelion can have a large impact on the temperature depending on the distance to the central object. In the cometary community, however, H/D exchange is often modeled only in the coma gas-phase chemistry via several reactions \citep{Rodgers:2002} following sublimation of ices.

The mechanisms underlying proton/deuteron scrambling have already been studied extensively in the 1980's in the physical chemical community, in particular by Devlin and co-workers \citep{Thornton:1981, Bertie:1983, Collier:1984, Woolridge:1988} who studied H/D exchange in water ices as well as water-ammonia mixtures with different isotopic compositions. For the case of water ice, exchange has been found to occur both in the bulk \citep{Collier:1984} and on the surface \citep{Uras:2001, Park:2004, Moon:2010}. In the case of isolated \ce{D2O} molecules in a \ce{H2O} environment, the prevailing mechanism is often referred to as the so-called `hop-and-turn' mechanism. In Fig.~\ref{waterstruct}, three scenarios of the local ice structure are depicted: (a) a perfect crystalline structure, (b) an ionic defect, \emph{i.e.}, the presence of \ce{H3O+}, and (c) an L-defect, \emph{i.e.}, the lack of an expected hydrogen bond. Under influence of an ionic defect, the heavy water is converted into two coupled HDO entities that share a hydrogen bond. These can be converted into two nearest-neighbour HDO molecules by passage of an L defect. Finally, two isolated molecules are created after an additional proton transfer. Such a detailed mechanism, with corresponding distinct infrared (IR) spectra (in the O-D stretching region) cannot be resolved for current astronomical IR spectra. For astrochemical purposes, the overall reaction can therefore be summarized as 
\begin{align}
\ce{H2O} + \ce{D2O} &\xrightarrow{k_\text{f}} \ce{2 HDO} \;\; \tag{R1}\label{R1} \\
\ce{2 HDO} &\xrightarrow{k_\text{r}} \ce{D2O} + \ce{H2O}. \;\; \tag{R2}  \label{R2}
\end{align}

\citet{Collier:1984} and \citet{Woolridge:1988} studied cubic ices with \ce{D2O} concentrations of several percent in a high vacuum setup. The activation energy for proton exchange was found to be $\sim$5000~K, which means that the process is not likely to be relevant on interstellar time scales at grain temperatures up to the sublimation temperature of ice at 90-100~K. Since the proposed hop-and-turn mechanism depends on the existence of point defects present in the ice and on the defect concentrations, the study was extended to amorphous water samples by \citet{Fisher:1995}. Indeed, they found a much lower activation energy of the turn step involved, $\sim$3000~K. Studies of exchange on ice surfaces -- also subject to larger structural defects -- indicated an enhanced L-defect activity \citep{Uras:2001}. These studies were performed, however, using ices doped with \ce{HCl} in order to have a higher sensitivity. Protons are generated via the reaction {\ce{HCl + H2O -> H3O+ + Cl-}}, the exothermicity of which could influence the local processes that occur \citep{Kim:2009}. Doping creates additional (shallow) proton traps in the ice in the vicinity of the counterion \citep{Uras:2001}, and proton films can remain inactive up to around 125~K \citep{Lee:2007}, both of which can influence the reaction rates detected. \citet{Moon:2010} showed that doping lowers the activation energy found for H/D exchange on surfaces by almost a factor 2. 

A similar situation applies to space. In the ISM ices do not have a perfectly ordered structure, partially because they are (i) not condensed at high enough temperatures, (ii) comprised of many species that are formed on the surface, (iii) composed of more molecules than only water, and (iv) subject to structural changes upon energetic processing of the ice. These effects will contribute to a high defect concentration in interstellar ices. 

In this paper, we aim to bring together the previous results of these scientific communities -- physical chemistry, cometary chemistry and (laboratory) astrochemistry -- while presenting several new experimental results of which the relevance to astronomy is discussed. We discuss amorphous \ce{D2O}:\ce{H2O} ices with mixing ratios around 1:1. Although such ratios are not astrochemically relevant, they do allow for a higher sensitivity of the proton exchange process at lower temperatures, without the neccessity to use doping.  We mimic the high expected amount of defects in the ISM by intentionally growing amorphous ice structures at temperatures of 15~K and probe whether this higher defect concentration in the ice structure, see Fig.~\ref{waterstruct}, leads to an increase of the reaction rate for reaction~\ref{R1}. The results are extended to a more general concept taking into account `all' hydrogen bonded molecules. Subsequently, the astrochemical implications are discussed focussing on the relevance of proton/deuteron swapping at long time scales, but at temperatures below full ice desorption.

\section{Methods}
All experiments studied here consist of two sequential steps: (a) the simultaneous deposition of \ce{H2O} (Milli-Q) and \ce{D2O} (Sigma-Aldrich 99.96\%) on the substrate at 15~K and subsequently (b) an isothermal experiment at a given temperature ($T_\text{iso} \geq 90$~K) during which the level of proton exchange is probed by means of reflection absorption infrared spectroscopy (RAIRS). In one experiment (Section~\ref{OtherMol}) also a quadrupole mass spectrometer (QMS) is used. Below, the experimental procedure is described first, followed by an explanation of the calibration experiments performed, the analysis of the spectra, and the analysis of the temporal evolution of the surface abundances of \ce{H2O}, \ce{HDO} and \ce{D2O}.

\subsection{Experimental}
\begin{table}
\caption{List of (calibration) experiments and corresponding parameters; temperature ($T$), beamline angle to the plane of the surface, and molecular flux ($f$).}\label{exps}
\begin{tabular}{@{}l@{ }llllll@{}}
\hline
\multicolumn{7}{c}{Experiments}\\
\hline
	& $T_{\textrm{iso}}$& 90$^{\circ}$  & $f_{\textrm{dep}}$ 	& 135$^{\circ}$  & $f_{\textrm{dep}}$ 	& $t_{\textrm{iso}}$ \\
	& (K)	&		& (cm$^2$ s$^{-1}$)		& 			&  (cm$^2$ s$^{-1}$)		& (min) \\
\hline
1	& 90	& \ce{D2O}	& 10 $^{(12)}$		& \ce{H2O}		& 10 $^{(12)}$	& 270 \\
2	& 100	& \ce{D2O}	& 10 $^{(12)}$		& \ce{H2O}		& 10 $^{(12)}$	& 270 \\
3	& 120	& \ce{D2O}	& 10 $^{(12)}$		& \ce{H2O}		& 10 $^{(12)}$	& 360 \\
4	& 125	& \ce{D2O}	& 10 $^{(12)}$		& \ce{H2O}		& 10 $^{(12)}$	& 270 \\
5	& 130	& \ce{D2O}	& 10 $^{(12)}$		& \ce{H2O}		& 10 $^{(12)}$	& 210 \\
6	& 130	& \ce{D2O}	& 4 $^{(12)}$		& \ce{H2O}		& 10 $^{(12)}$	& 190 \\
7	& 135	& \ce{D2O}	& 10 $^{(12)}$		& \ce{H2O}		& 10 $^{(12)}$	& 150 \\
8	& 140	& \ce{D2O}	& 10 $^{(12)}$		& \ce{H2O}		& 10 $^{(12)}$	& 110 \\
9	& 140	& \ce{D2O}	& 4 $^{(12)}$		& \ce{H2O}		& 10 $^{(12)}$	& 120 \\
10	& 140	& \ce{H2O}	& 4 $^{(12)}$		& \ce{D2O}		& 10 $^{(12)}$		& 120 \\
\hline
11	& 130	& \ce{D2O}	& 10 $^{(12)}$		& \ce{NH3} 		& 7 $^{(12)}$		& 120 \\
12 $^{a}$& TPD	& \ce{D2O}	& -			& \ce{NO + H}		& -		& - \\
\hline
\multicolumn{7}{c}{Calibration experiments}\\
\hline
1	& var.  $^{b}$ & \ce{H2O}		& 20 $^{(12)}$	& 	-	& -	 & -\\
2	& var.  $^{b}$ & HDO		& 10 $^{(12)}$	$^{c}$	& 	-	& -	 & -\\
3	& var.  $^{b}$ & \ce{D2O}	& 19 $^{(12)}$ $^{d}$ & 	-	& -	 & -\\
4	& var.  $^{b}$ & -		& -			& -		& -	& -\\
\hline
\end{tabular}
Note. The notation $\alpha$ $^{(\beta)}$ implies $\alpha \times 10^{\beta}$ .\\
$^{a}${Instead of a full co-deposition, a layered experiment was performed, first co-depositing \ce{NO + H} to prepare a \ce{NH2OH} layer, which was then covered by \ce{D2O}.}
$^{b}${Spectra were acquired at all temperatures relevant for the experiments, \emph{i.e.} 90-140~K. }
$^{c}${Due to the mixture preparation of HDO, the statistical ratio between the constituents \ce{H2O}:HDO:\ce{D2O} is 1:2:1.} 
$^{d}${The effective deposition rate is somewhat lower as a result of HDO contamination in the \ce{D2O} sample, this is estimated to be $\leq 5\%$ from the OH stretching mode.} 
\end{table}

Experiments are performed using the SURFRESIDE$^2$ setup, which has been constructed to systematically investigate solid-state reactions leading to the formation of molecules of astrophysical interest at cryogenic temperatures. The setup has been extensively described in \citet{Ioppolo:2013} and therefore only a brief description of the procedure is given here. All performed experiments and the corresponding molecular deposition rates are listed in Table~\ref{exps}. The results of experiments 1-10 are discussed in Section~\ref{RnD}. Experiment 11 follows a procedure similar to that for the water experiments described below and is discussed in Section~\ref{OtherMol}. The method used for experiment 12 deviates and is discussed in Section~\ref{OtherMol}.

SURFRESIDE$^2$ consists of three UHV chambers: one main chamber for ice growth, ice processing, and ice diagnostics, and two chambers comprising atom beam lines. All have a room-temperature base-pressure between $10^{-9}-10^{-10}$ mbar. A rotatable gold-coated copper substrate in the center of the main chamber is cooled to the desired temperature using a He closed-cycle cryostat with an absolute temperature accuracy of $\leq$ 2~K. Both water and heavy water are prepared in a separate pre-pumped ($\leq 10^{-5}$ mbar) dosing line and after several freeze-pump-thaw cycles a co-deposition of room-temperature vapor \ce{H2O} and \ce{D2O} is performed. One species is deposited through a metal deposition line under an angle of 90$^{\circ}$ and the other through one of the separate UHV beam lines at an angle of 135$^{\circ}$ with respect to the plane of the surface, see Table~\ref{exps}. This UHV beam line can be operated independently and can be separated from the main chamber by a metal shutter. Deposition takes place at a surface temperature of 15~K to ensure a large amount of amorphicity of the ice (see discussion above). 

\begin{figure}
\begin{center}
\includegraphics[width=80mm]{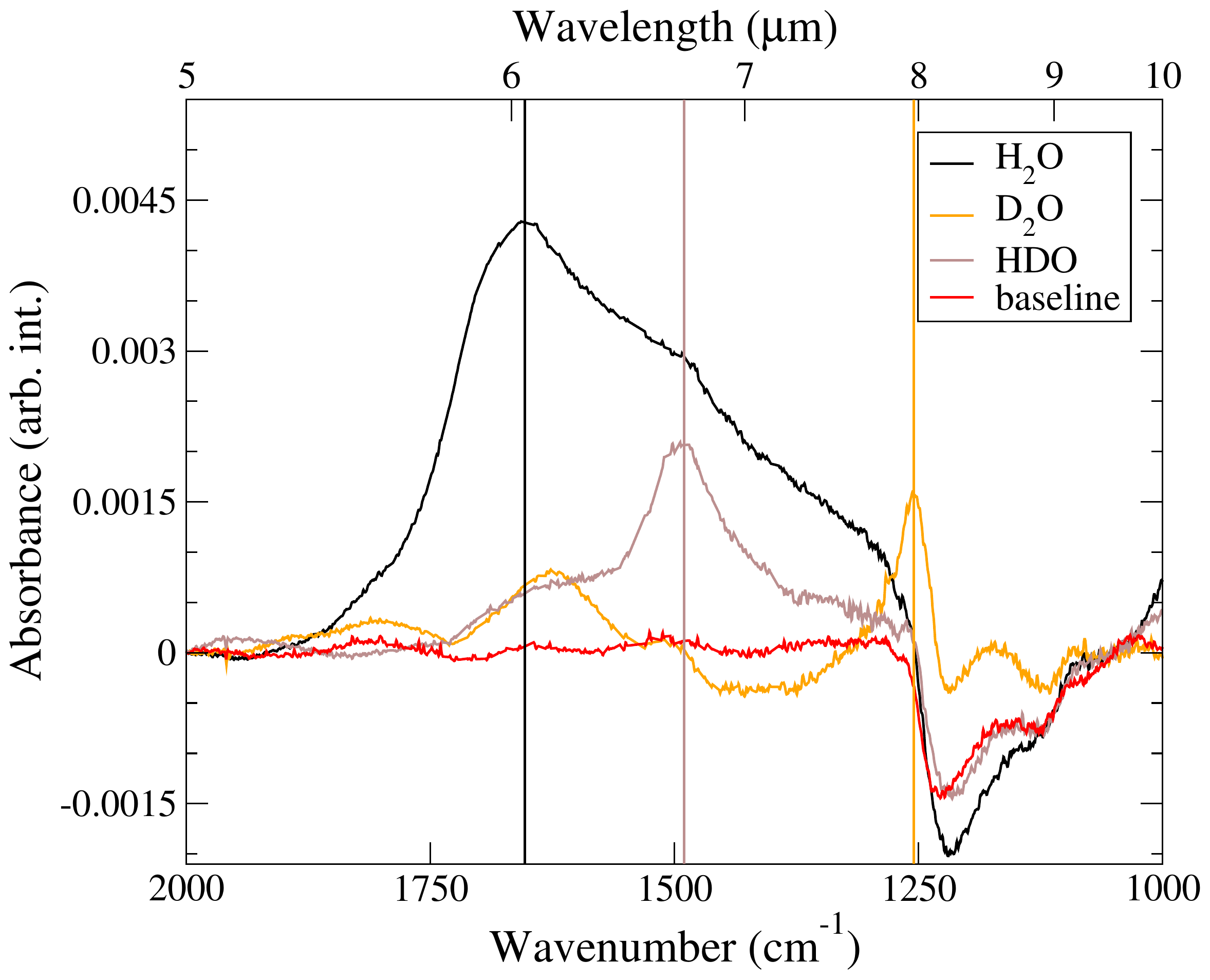}
\end{center}
\caption{Bending region of the calibration spectra used for fitting the experiments, all spectra are recorded at 130~K. See text concerning the 1625 cm$^{-1}$ band in the \ce{D2O} spectrum, the contamination in the HDO band, and the baseline artifact around 1200 cm$^{-1}$. }
\label{pure}
\end{figure}

After approximately 60 minutes of co-deposition at low temperature, yielding 45-65~ML, the substrate is heated up to the desired temperature, $T_\textrm{iso}$, between 90 and 140~K. Note that in the laboratory ices sublimate not at 100~K, but rather between 145-165~K, hence the higher temperatures employed here. The warm-up phase clearly affects the porosity of the ice \citep{Bossa:2012} and therefore also the ice structure. However, using \ce{H2O}:\ce{D2O} mixing ratios around 1:1, each molecule initially has at least one neighbor of its isotopic counterpart, which renders the degree of porosity less important. Note also that the extra collapse of the pores at 120~K with respect to that at 90~K is only 3\% \citep{Bossa:2012}. Thus, assuming locally distorted hydrogen bonded structures \citep{Karssemeijer:2014, Karssemeijer:2014pc}, the exact structure of the ice and/or diffusion mechanisms \citep{Jung:2004, Oxley:2006} do not play a role here.
A RAIR difference spectrum with respect to the background is acquired every 5 or 10 minutes up to the final time of the experiments, $t_\textrm{iso}$ in Table~\ref{exps}. The background spectrum is acquired from an empty surface prior to the co-deposition at low temperature. RAIR spectra comprise a spectral range between 4000 and 700 cm$^{-1}$ with a spectral resolution of 1 cm$^{-1}$ and are averaged over 512 scans. Exchange occurs faster at higher temperatures, therefore those experiments have a shorter duration. Pure component calibration spectra are used for the analysis of the experiments, see Section~\ref{SpecFit}. 

\begin{figure}
\begin{center}
\includegraphics[width=80mm]{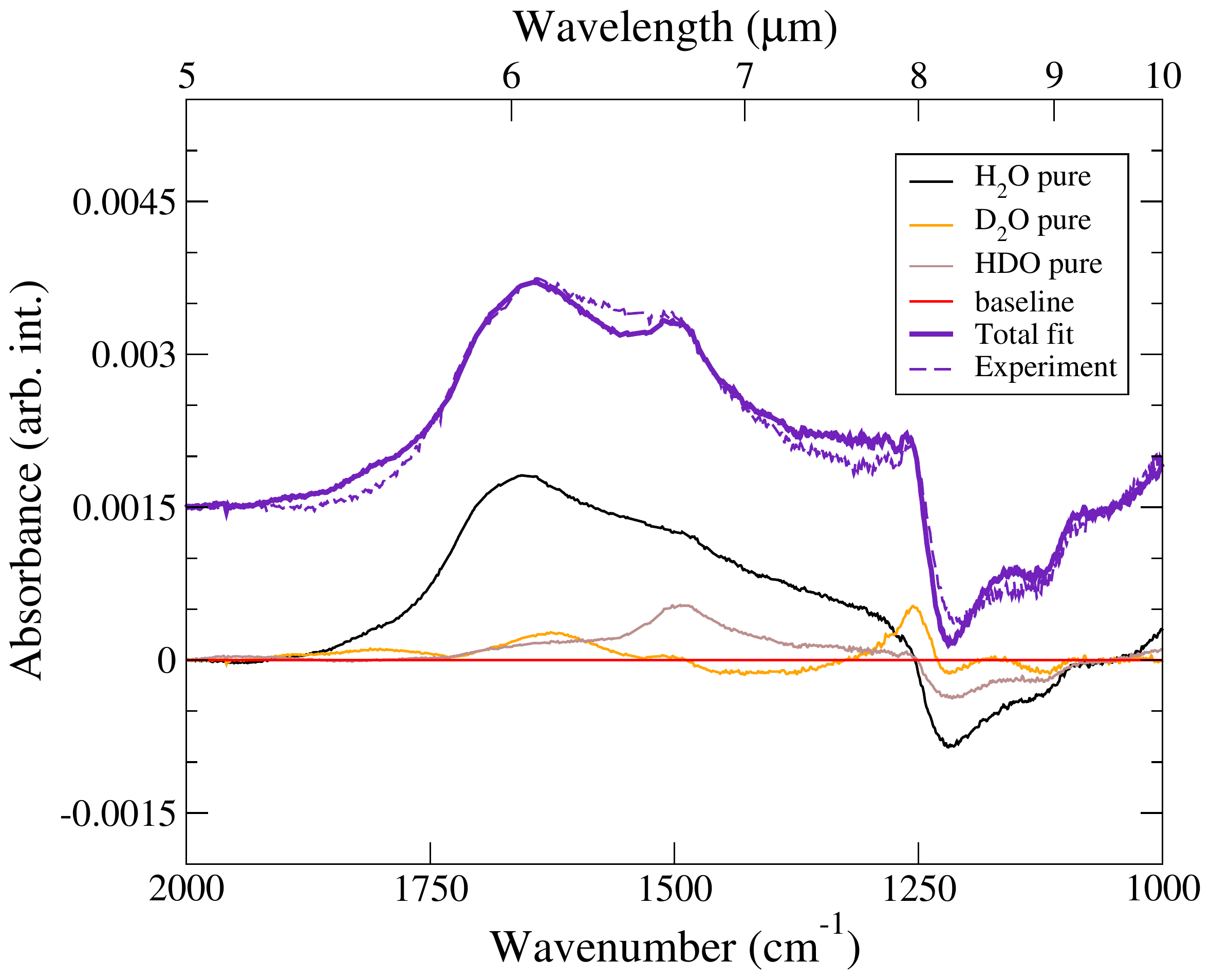}
\end{center}
\caption{Example of a recorded spectrum in the water bending region (130~K) and its best fit, decomposed into the separate components (\ce{H2O}, HDO, and \ce{D2O} - see Fig.~\ref{pure}), recorded after 210 minutes. Note: this concerns experiment 5 from Table~\ref{exps}. Pure spectra themselves compensate for the baseline artifact at 1200 cm$^{-1}$.}
\label{example130K}
\end{figure}

The deposition rates mentioned in Table~\ref{exps} are calculated using the following relation 
\begin{equation}
 \frac{c_{\ce{X2O}} \cdot P_{\ce{X2O}}\cdot \langle v \rangle}{4\cdot k_{\rm B}\cdot  T}
\end{equation}
where $c_\ce{X2O}$ is the calibration factor for the pressure gauge for the three isotopologues of water (\ce{X2O} = \ce{H2O}, \ce{HDO}, or \ce{D2O}), $v$ is the thermal velocity of the vapor molecules at 300 K, $k_{\rm B}$ is the Boltzmann constant, and $T$ corresponds to the (room) temperature. The calibration factor for water is 1/0.9. Since there is no significant difference found between the absolute partial cross sections for electron-impact ionization of \ce{H2O} and \ce{D2O} (influencing the pressure reading of the gauge) \citep{Straub:1998, Itikawa:2005}, we assume that this is also the case for HDO and all calibration factors for the pressure are taken to be equal. 

Note that the initial spectra of experiments 1-5, 7 and 8 at 15~K after deposition are compared with each other to confirm the reproducibility: each low-temperature spectrum is analyzed with the fitting procedure described below and the initial surface abundances of water and deuterated water are found to be reproducible with a standard deviation of $\sim$ 5\%.

\begin{figure*}
\begin{center}
\includegraphics[width=155mm]{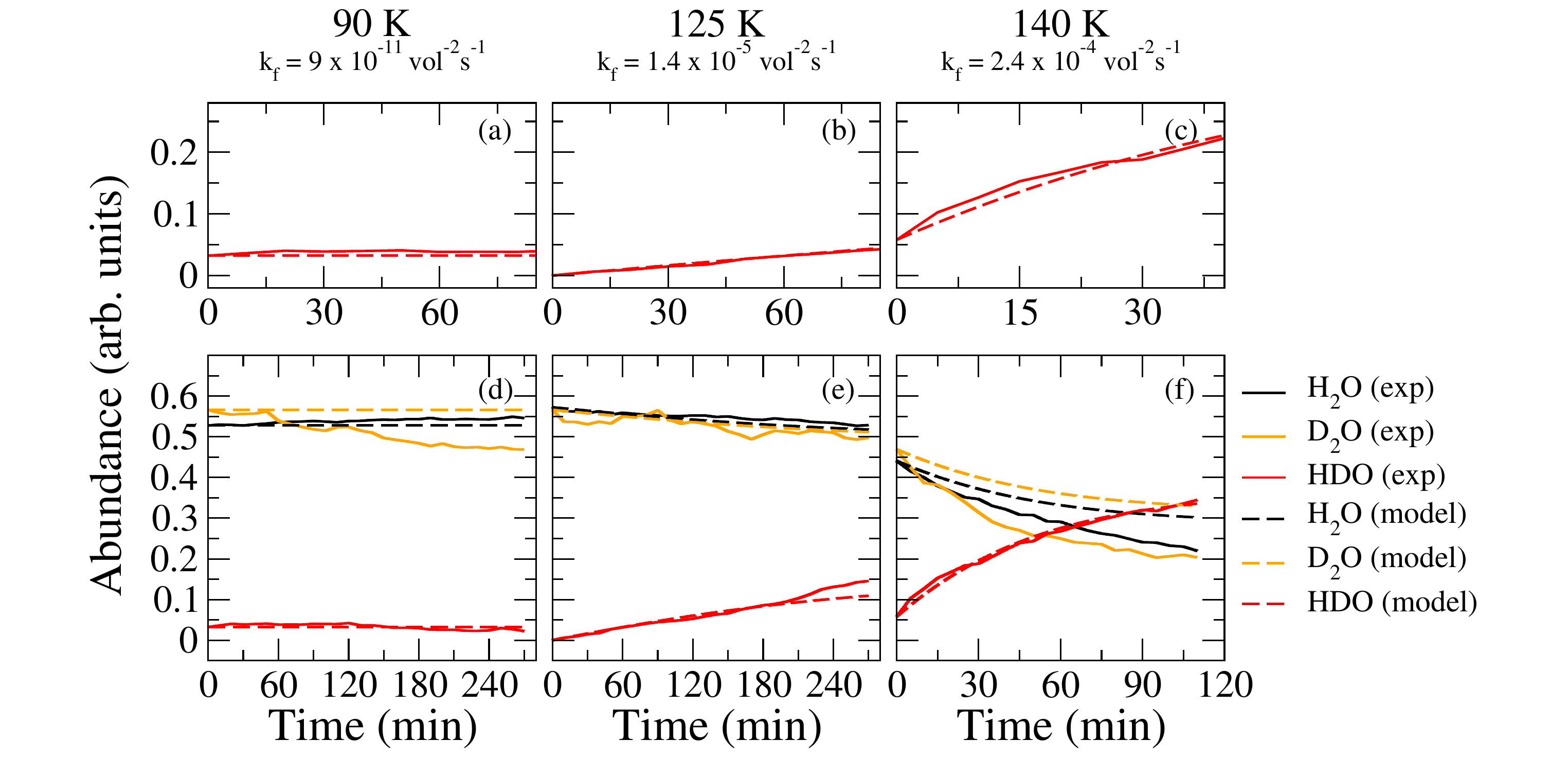}
\end{center}
\caption{Modeled \ce{H2O}, \ce{D2O}, and HDO concentrations for experiments 1 (a and d), 4 (b and e), and 8 (c and f) (Table~\ref{exps}). The top row, panels (a)-(c), are zoom-ins of the HDO experimental and modeled abundance for the first 33\% of the experimental duration. The bottom row, panels (d)-(f), depicts the evolution of all three components for the entire duration.}
\label{fits3T}
\end{figure*}

\subsection{Spectral Fitting}\label{SpecFit}
Each spectrum is the sum of the pure \ce{H2O}, HDO, and \ce{D2O} components along with their intermolecular interactions. The intermolecular interactions cannot be captured in pure calibration spectra and they are stronger in the IR stretching than bending region. Futhermore the bending region is less sensitive to crystallization effects. Therefore, our region of interest lies in the 2000-1000 cm$^{-1}$ range, where the \ce{H2O}, HDO and \ce{D2O} molecules vibrate in their respective bending modes: 1660, 1490, 1250 cm$^{-1}$. Upon proton/deuteron exchange the \ce{H2O} and \ce{D2O} intensities decrease, whereas the HDO intensity increases. 

The integrated area of each mode can be converted into a number of molecules, assuming that band strengths are available, but in reflection mode these are typically ill-constrained. To be able to quantify the dynamics at work, a spectral fitting procedure is needed to separate the three components. For this purpose, also three calibration experiments are performed. 

A calibration experiment, see final rows of Table~\ref{exps}, consists of the deposition of a `pure' X$_2$O component for 30 minutes at 15~K and spectrum acquisition thereafter at all temperatures relevant for the experiments performed, \emph{i.e.}, 90, 100, 120, 125, 130, 135, and 140~K. At each temperature, the spectrum looks slightly different and therefore pure spectra for each temperature are necessary for the fitting procedure. Furthermore, in the spectral region between $\sim$1050 and $\sim$1250 cm$^{-1}$ a known temperature-dependent artifact is present, which makes it neccessary to record blank spectra at all relevant temperatures as well. {It is most likely due to a temperature dependence in the absorption of our background sample.} These spectra are also included in the fits (calibration experiment 4), but do not fully correct for the effect, as explained below. 

Using the pure components, we avoid having to correct for the difference in band strengths: each pure spectrum consists of a known amount of deposited \ce{H2O}, HDO, and \ce{D2O}. A second band around 1625 cm$^{-1}$ in the \ce{D2O} spectrum has its origin in the combination mode $\nu_\text{bending} + \nu_{\text{libration}}$ \citep{Bertie:1964}. There is also a small OH stretching signal present in the spectrum , but it is $\leq 5\%$ with respect to the OD stretching signal. HDO was prepared by mixing equal amounts of \ce{H2O} and \ce{D2O} vapors into a pre-pumped ($\leq 10^{-5}$ mbar) dosing line. We expect the hydrogen and deuterium atoms to be distributed statistically, resulting in a statistical 1:2:1 mixture. That means that the effective HDO deposition rate is only half of the total deposition rate for this particular experiment. The spectra shown in Fig.~\ref{pure} are used for the fitting procedure. The HDO spectrum cannot be obtained pure, due to the way it is produced in the dosing line. In the fitting procedure we used both the original spectrum as well as a spectrum corrected for the \ce{H2O} and \ce{D2O} contamination in the HDO contribution, see Section~\ref{OptPro}.

Every separate experiment is composed of a time-resolved series of recorded spectra, each providing a snapshot of the ongoing proton exchange process. A non-negative least squares solver is used to fit the pure components to the spectrum only in the bending range of 2000-1000 cm$^{-1}$. Figure~\ref{example130K} depicts an example of such a fit for the final spectrum (recorded after 210 min) recorded during experiment 5 from Table~\ref{exps}. In this case, the baseline does not contribute to the fit since the sum of the pure components themselves already compensate for the artifact. {Usually, however, the temperature-dependent baseline is needed to take care of this artifact. }

\subsection{Reaction dynamics}\label{ReactionDyn}
The total contribution of each pure component to the fit is subsequently plotted versus time in order to resolve the dynamics. We are not able to observe the isolated HDO entities in a water matrix that have been proposed to be produced through a hop-and-turn mechanism \citep{Collier:1984}, because both reaction partners are present in the ice with mixing ratios between 1:2 to 2:1. Because of this, our system is intrinsically easier to model, since we observe only the proton or deuteron hopping, leading to a reaction system of the forward and backward reactions, \ref{R1} with rate $k_\text{f}$ and \ref{R2} with rate $k_\text{r}$. They can be evaluated in terms of a mean field approximation. The temporal evolutions of the concentrations (or abundances) can be described with a set of ordinary differential (rate) equations 
\begin{align}
 \frac{\rm d[\ce{H2O}]}{\rm dt} &= -k_\text{f} [\ce{H2O}] [\ce{D2O}] + k_\text{r} [\ce{HDO}]^2 \label{kh2o}\\
 \frac{\rm d[\ce{D2O}]}{\rm dt} &= -k_\text{f} [\ce{H2O}] [\ce{D2O}] + k_\text{r} [\ce{HDO}]^2 \label{kd2o}\\
 \frac{\rm d[\ce{HDO}]}{\rm dt} &= 2\;k_\text{f} [\ce{H2O}] [\ce{D2O}] -2\;k_\text{r} [\ce{HDO}]^2 \label{khdo}\;.
\end{align}
Where $\frac{\rm d[\ce{H2O}]}{\rm dt} + \frac{\rm d[\ce{D2O}]}{\rm dt} + \frac{\rm d[\ce{HDO}]}{\rm dt} = 0$. {We assume thermal desorption to be negligable at these temperatures.} Using an optimization algorithm in conjuction with a least-square ODE solver, $k_\text{f}$ and $k_\text{r}$ can be extracted for each experiment. $[\ce{H2O}]_0$, $[\ce{HDO}]_0$, and $[\ce{D2O}]_0$ are taken equal to the values found for each experiment. Section~\ref{OptPro} discusses several examples of both the temporal evolution of the pure components as well as the modeled result via the rate equations. Note that the start of each experiment is almost exclusively sensitive to $k_\text{f}$, since $[\ce{HDO}]_0$ is always low. We therefore chose to use the first 33\% in time, while still in the linear regime, of the experiment to determine $k_\text{f}$. A discussion on this topic is given in the section below.

\subsection{Optimization procedure}\label{OptPro}
Figure~\ref{fits3T} depicts three examples of our modeled results, for temperatures of 90, 125, and 140~K (experiments 1, 4, and 8 in Table~\ref{exps}). The top row, panels (a)-(c), shows the HDO fit to the experiments for the first 33\% of the experiment, and from the (near-) linearity of the experimental slopes it can be deduced that it is indeed sensitive mainly to $k_\text{f}$. In panels (d)-(f) the time evolution of the experimental and modeled abundances (Eqns.~\ref{kh2o}-\ref{khdo}) for all three species is plotted for the entire experimental time.

Analyzing the abundances at 90~K we find an apparent discrepancy for the temporal evolution of \ce{D2O}. While the recorded \ce{H2O} and HDO abundances remain constant in time, since there is no proton exchange at this low temperature, the \ce{D2O} abundance does not. Inspection of the IR spectra shows that this is likely caused by the spectroscopic artifact between $\sim$1050 and $\sim$1250 cm$^{-1}$ mentioned above. This artifact actually partially overlaps with the \ce{D2O} signal. We therefore decided not to include the heavy water abundance explicitly in the determination of the optimal reaction rates, but to check in retrospect whether or not the modeled abundance is in approximate agreement with the experiments. 

{Furthermore, the graphs at high temperature (140~K) show that both modeled \ce{H2O} and \ce{D2O} abundances deviate somewhat from the experiment. To further constrain this, we have studied two limiting cases of the dependence of the \ce{H2O} and \ce{D2O} abundances on the HDO calibration spectrum: (a) assuming a `pure' HDO calibration spectrum and (b) assuming a statistical 1:2:1 distribution between \ce{H2O}:HDO:\ce{D2O} for the deposited HDO. Unfortunately, the HDO ice which we deposited to obtain our HDO calibration spectrum did not only contain HDO, but also \ce{H2O} and \ce{D2O}. If we do not account for this in our calibration spectrum (case (a)), the water and heavy water contribution are underestimated during the fitting procedure. Accounting for the presence of both contaminants in the HDO calibration spectrum by assuming a statistical distribution (case (b)) will on the contrary lead to an overestimation of the \ce{H2O} and \ce{D2O} abundances. Therefore, we decided to optimize the model and obtain $k_\text{f}$ and $k_\text{r}$ by using  only the HDO abundance. Subsequently, we verified that the modeled water and heavy water abundances fall within the range determined by the two limiting cases for the obtained values of  $k_\text{f}$ and $k_\text{r}$.}

Thus, the optimization sequence used for modeling the HDO abundance is as follows: 
\begin{enumerate}
\item $k_\text{f}$ is optimized by modeling the first 33\% in time of the experiment with $k_\text{r}$ = 0
\item $k_\text{r}$ is optimized by modeling the final 33\% in time of the experiment with $k_\text{f}$ fixed to the value found in (i)
\item $k_\text{f}$ is optimized again by modeling the full experiment with $k_\text{r}$ fixed to the value found in (ii).
\end{enumerate}
The error function used for this optimization is
\begin{equation}
E_\text{HDO} =  \sum_{t_1}^{t_2} \frac{1}{\Delta t} \; \frac{ \left ( [\ce{HDO}]_\text{model}(t) - [\ce{HDO}]_\text{exp}(t) \right ) ^2 }{ \left< [\ce{HDO}]_\text{model}(t) + [\ce{HDO}]_\text{exp}(t) \right>  } \;.\label{error}
\end{equation}
The reaction rates leading to the lowest error are selected to be used in next iterations or are stored as the optimum rates.

Finally, the activation energy of \ref{R1} is determined by an Arrhenius fit of the different reaction rates versus temperature. The error associated with this fit is determined with the stats subpackage of scipy in python and corresponds to the standard error of the slope.

\section{Results and Discussion}

\subsection{Activation energy of proton exchange in \ce{H2O}:\ce{D2O} mixtures}\label{RnD}
\begin{figure}
\begin{center}
\includegraphics[width=80mm]{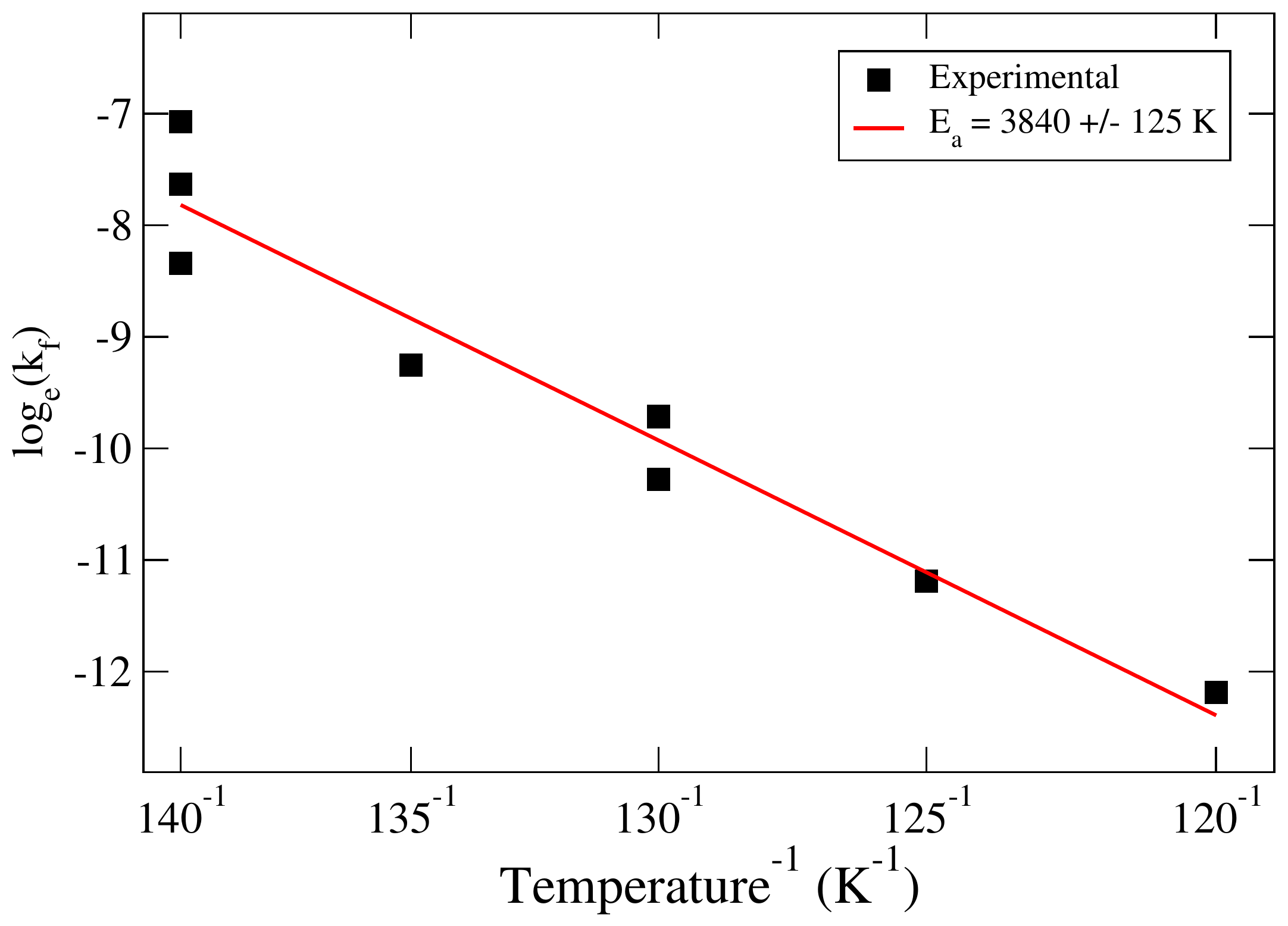}
\end{center}
\caption{Arrhenius plot for $k_\text{f}$. }
\label{arrhenius}
\end{figure}

The reaction rates obtained from the reaction dynamical fitting by a simple rate equation model are converted into an activation energy with the use of an Arrhenius expression: 
\begin{equation}
 k = \nu \;\exp\left(\frac{-\text{E}_\text{a}}{k_\text{B}\,T}\right). \label{eqn_arrh}
\end{equation}
In this expression the prefactor, $\nu$, represents a trial frequency related to the vibrations of a species in a local potential well and corresponds to values between $10^{12}-10^{13}$ s$^{-1}$ \citep{Hasegawa:1992}.
Figure~\ref{arrhenius} depicts ln($k_\text{f}$) versus 1/$T$. If experiments 2-10 are included in the determination of the activation energy, a value of $3840 \pm 125$~K is found. The error given here is the error of the slope. The exact value found for the activation energy changes when the amount of assumed contamination in the HDO spectrum is varied, \emph{i.e.}, going from case (a) to (b). The resulting values, however, are not significantly different as assessed by a two-sided student's t-test ($p<0.05$). 

As stated before, \citet{Woolridge:1988} found a substantially higher activation energy of $\sim$5000~K associated with reaction \ref{R1}. They proposed a mechanism of proton transfer that occurs via ionic defects in the ice. These defects will be more prominent in an amorphous ice and indeed the rate increases drastically; $e^{-3840/T}/e^{-5000/T} > 4000$ for $T \leq 140$~K. The specific implications of this for astrochemical environments are discussed in Section~\ref{astroim}.

Experiments below 130~K are not sensitive to $k_\text{r}$, because the situation in which enough HDO molecules are next to each other is never reached within the experimental duration. For our experiments at 130-140~K it is not possible to find any correlation between $k_\text{r}$ and the temperature, again because for the largest part of the experiment there are not enough HDO molecules close to each other. However, \citet{Collier:1984} suggested that $k_\text{r} < k_\text{f}$ in the case of an isolated \ce{D2O} molecule in an \ce{H2O} surrounding, as a result of the local structure. This can also be explained from a more theoretical free energy point-of-view. The total Gibbs free energy of the reaction, $G$, is determined by the enthalpies, $H$, and entropies, $S$, of the molecules:
\begin{equation}
 \Delta G (\text{\small 2\:\ce{HDO}-\ce{H2O}-\ce{D2O}}) = \Delta H_\text{f} - \text{T}\cdot \Delta S_\text{f}
\;.
\end{equation}
Using the gas-phase data reviewed in \citet{Chase:1998} this leads to the following expression
\begin{align}
 \Delta G &= 290 - T\cdot 11.86 \;\;\;\text{ J mol}^{-1} \\ 
& = 34.88 - T\cdot 1.43 \;\;\text{ K}\;.
\end{align}
Although the thermochemistry in the solid state is affected by the local structure, the gas-phase expression given above does show that the driving force is the entropy change and that the reaction becomes clearly {exogonic} at higher temperatures only, hence we also expect that $k_\text{r} < k_\text{f}$. If this relation is incorporated into the optimization procedure, the activation energy found does not differ significantly from 3840~K. 

Finally, note that here no conclusions can be drawn concerning the prefactor $\nu$, because accurate values for the band strengths are lacking. 

\subsection{Proton exchange in other hydrogen bonded molecules}\label{OtherMol}
As mentioned in the Introduction, proton exchange in ices has been studied previously for several (mixtures of) hydrogen-bonded molecules. Here, we mention these studies specifically and extend the studies of water to a more general ice theme: proton exchange in hydrogen bonded molecules. As long as so-called proton wires exist between the hydrogen bonded molecules, exchange is expected to be rapid \citep{Bertie:1983}. A proton wire can be thought of as a chain of hydrogen bonded molecules that has the flexibility to transfer protons from one to the next molecule. In such a way, any ionic defect in the ice can be easily passed on to the neighboring molecule and hence protons and deuterons can move efficiently. The following ice mixtures are briefly discussed: \ce{NH3}:\ce{D2O}, \ce{CD3OD}:\ce{H2O}, \ce{CD3ND2}:\ce{H2O}, and \ce{NH2OH}:\ce{D2O}.

Exchange between isolated \ce{NH3} in an amorphous \ce{D2O} ice was found to be faster than in pure water ice and much faster than for crystalline ammonia \citep{Thornton:1981, Bertie:1983}. We have performed a similar experiment, co-depositing \ce{NH3} and \ce{D2O} (2:3) at low temperature and heating to 130~K, experiment 11 in Table~\ref{exps}. Indeed rapid exchange of protons can be confirmed. A steady-state HDO concentration is reached at least three times faster than in the case of water (experiment 5). This shows once more that amorphicity can enhance the exchange rates.

Concerning the case of isolated amorphous d4-methanol mixed with water both \citet{Ratajczak:2009} and \citet{Souda:2003} confirmed thermal exchange to be efficient at laboratory time scales starting from 120~K. Note that this exchange concerns specifically the hydroxyl group of methanol. The C-D (or C-H) bonds do not participate in any hydrogen bonding network and are therefore not exchanged via this mechanism. They can, however, be exchanged via deuterium bombardment of \ce{CH3OH} as discussed by \citet{Nagaoka:2005}.

\citet{Ratajczak:2012} also studied the exchange of isolated methylamine (\ce{CD3ND2}) with water and found similar results as for methanol: exchange takes place only on the amine entity of the molecule and is detectable from 110~K onwards.

\begin{figure}
\begin{center}
\includegraphics[width=85mm]{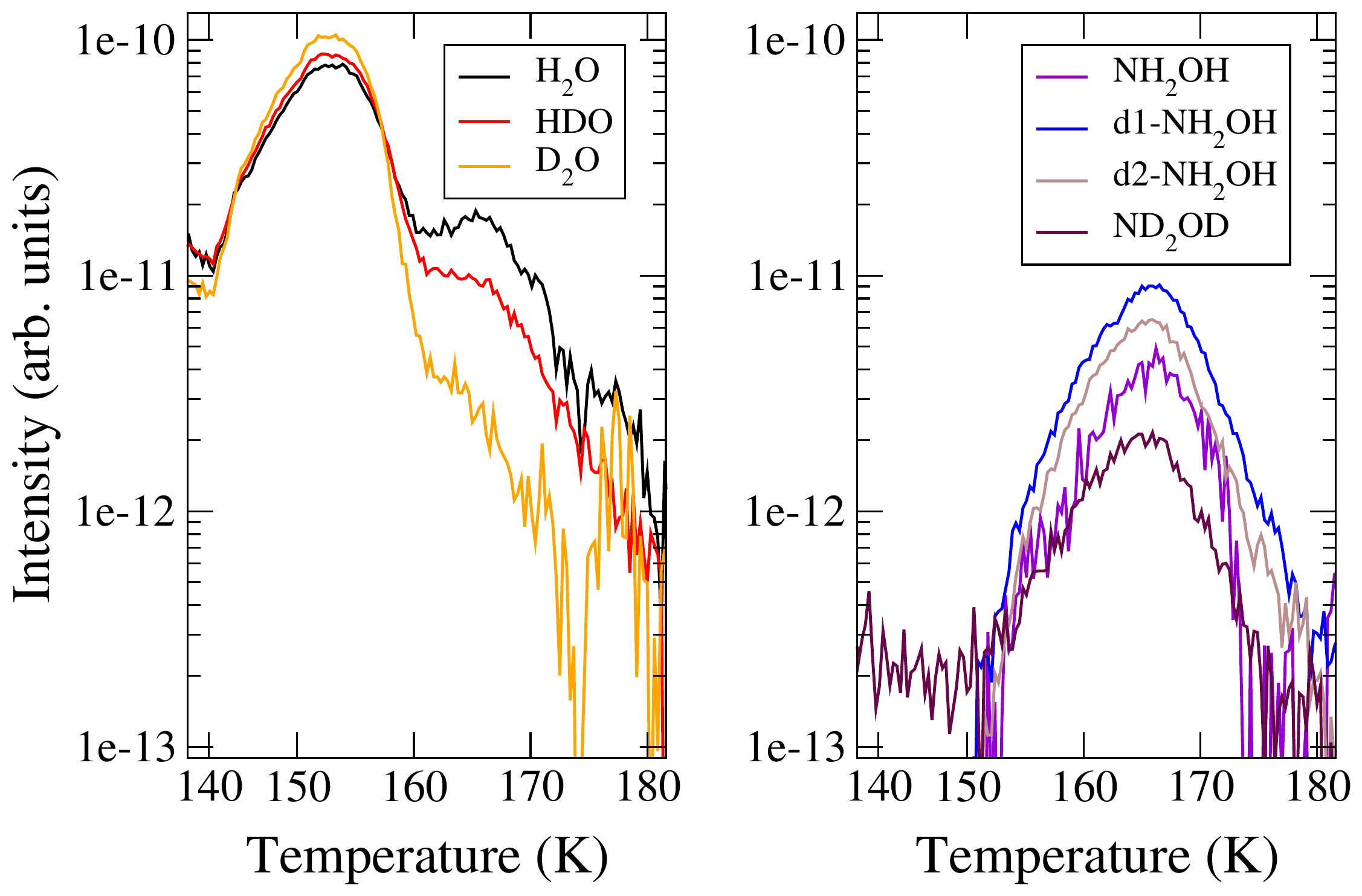}
\end{center}
\caption{{QMS traces for TPD experiments of mass to charge ratios of 18, 19, 20, 33, 34, 35, and 36 probing exchange in a layered \ce{NH2OH}:\ce{D2O} ice. Note that the x-axis changed to temperature.}}
\label{QMS}
\end{figure}

Here, in light of recent studies highlighting the hydrogenation of NO molecules and the subsequent formation of \ce{NH2OH} \citep{Congiu:2012, Fedoseev:2012}, we also performed an experiment probing the exchange between \ce{D2O} and \ce{NH2OH} (Experiment 12, Table~\ref{exps}). In this case both hydrogen bearing groups are able to form hydrogen bonds. A layer of \ce{NH2OH} is grown following a similar procedure as described by \citet{Fedoseev:2012} and capped with a layer of \ce{D2O} at 15~K. Increasing the sample temperature to 130 or 140~K could lead to exchange of protons and deuterons on the interface of the two molecular layers. However, proton exchange is not clearly visible by RAIR spectroscopy due to overlapping band frequencies between the NH(D) and OD(H) groups. Therefore, a temperature programmed desorption experiment is performed, during which the molecules gain more mobility increasing the probability that \ce{NH2OH} and \ce{D2O} meet, albeit for a short time until full desorption takes place. This experiment resulted in the detection of mass to charge ratios of 19, 34, 35, and 36 in the QMS, see Fig~\ref{QMS}. {In this Figure, the intensity does not reflect the in situ infrared intensity, but the intensity of gas-phase species after evaporation from the surface during a linear heating of the substrate. The temperature indicates the surface temperature at which this evaporation occurs.} These masses correspond to the molecules HDO, singly and doubly deuterated \ce{NH2OH}, and \ce{ND2OD}. This confirms proton exchange in the amine and hydroxyl groups, both of which have hydrogen bonds.

\section{Experimental Conclusions}
\noindent {The thermal process of proton/deuteron exchange between water and its isotopologues has been studied in mixed ices. Using RAIR spectra the time evolution of the characteristic vibrational bending modes of \ce{H2O}, \ce{HDO}, and \ce{D2O} has been monitered as a function of temperature. In particular, \ce{H2O} and \ce{D2O} have been grown at 15~K and heated up to a desired temperature where the occurence of HDO has been probed indicating that the reaction \ce{H2O + D2O -> 2 HDO} takes place. Temperatures between 90 and 140~K have been considered and a reaction rate for each temperature has been determined with the use of a simple rate equation model. The linearity of the resulting Arrhenius plot in Fig.~\ref{arrhenius} shows that it is reasonable to express the rate through thermal activation and an activation energy of 3840~K has been found. }

{ \citet{Ratajczak:2012} suggests that proton exchange coincides with crystallization. However, exchange has been found to take place prior and post crystallization, both in the work presented here as well as by \citet{Woolridge:1988, Fisher:1995, Galvez:2011}. In the ISM, (F)UV radiation would render the ices largely compact amorphous by destruction of local molecules and recombination of its fragments. Exothermic energy release upon molecule formation is likely to have the same effect \citep{Palumbo:2006, Oba:2009}.  }

{In this study, we mimic the lack of long-range order in interstellar ices by intentionally growing amorphous ice structures. The aim is to investigate whether a higher concentration of structural defects (see Fig.~\ref{waterstruct}) in amorphous ices with respect to crystalline structures leads to exchange rates high enough to become relevant on interstellar time scales. We indeed find a lower overall (or rate-determining) activation energy associated with the exchange in non-doped amorphous ices with non-energetic detection methods compared to crystalline ices: 3840~versus 5000~K.  This corresponds to the quantitative findings of \citet{Fisher:1995} for doped ices, those of \citet{Moon:2010} for pure ice surfaces and to the qualitative results mentioned by \citet{Galvez:2011}. The latter paper also discusses H/D exchange in water ices, but from the point of view to investigate whether HDO detections in ice are possible.}

{Assuming a value of $10^{12}$ s$^{-1}$ for the prefactor $\nu$ in Eqn.~\ref{eqn_arrh}, this corresponds to typical timescales as summarized in Table~\ref{timescale}. The large differences covering orders of magnitude of the typical timescales of proton exchange clearly show the importance of incorporating the correct activation energy.}

\begin{table}
 \caption{Typical thermal exchange timescales at various grain temperatures and activation energies.}\label{timescale}
\begin{tabular}{lll}
\hline
 $T$ (K) & $E_\text{a}$ = 5000~K $^{(1)}$ & $E_\text{a}$ = 3840~K $^{(2)}$\\
\hline 
 70	& $3.3\times10^{11}$ yr	& $2.1\times10^{4} $ yr	\\
 80 	& $4.4\times10^7$ yr	& $2.2\times10^{1}$ yr	\\
 90	& $4.2\times10^4$ yr	& $1.1\times10^{-1} $ yr	\\
\hline
\end{tabular}\\
$^{(1)}$ \citet{Woolridge:1988} $^{(2)}$ This work
\end{table}

\section{Astrochemical Implications}\label{astroim}
Interstellar ices are likely to be defect-rich by nature because of the many ice components and various types of (energetic) processing. Furthermore, since ices consist mainly of water, the hop-and-turn mechanism is likely to dominate proton exchange, while, simultaneously, the existence of the proton wires mentioned above is key to efficient transfer.

\subsection{Protostellar and protoplanetary environments}
The typical timescales derived with the experimentally found activation energy can be compared to interstellar timescales, \emph{i.e.}, those mentioned in Table~\ref{timescale} and those obtained for protostellar and protoplanetary disk environments. \citet{Schoier:2002} showed that the transit time for grains and molecules through the warm, dense region around the hot core IRAS 16293-2422 is of the order of several hundred years. Ices present at temperatures above 80~K can thus be influenced by scrambling. Furthermore, although the high dust-temperature regions in disks concern only the inner few mid-plane AU \citep{Walsh:2010}, turbulent vertical and radial mixing can result in transportation of water from the midplane to disk surface \citep{Furuya:2013, Albertsson:2014}. This is part of a cycle in which atomic oxygen is transported to the midplane and reforms water and/or other molecules. Since the grains are transported, also the time they pass at higher temperatures is longer, hence allowing proton scrambling to take place. This then applies to all hydrogen bonded molecules. Mixing timescales are determined by the ratio between the column density of water and the flux in upward or radial direction \citep{Furuya:2013}. They find that typical timescales range between $10^{4}$ and $10^{7}$ yrs for radii between 1 and 200 AU and that the radial accretion timescale is of a similar order of magnitude and are long enough to allow scrambling to take place.
\begin{itemize}
\item[] \emph{Comparing typical dynamical timescales in protostellar and protoplanetary disk environments to the values listed in Table~\ref{timescale}, we find that an activation energy of 3840~K renders the thermally activated H/D exchange relevant at static dust temperatures of 70~K and above.}
\item[] \emph{We would therefore expect a \ce{D2O} detection probability lower than determined statistically in high-temperature regions, because any \ce{D2O} available in the ice is likely to be converted into two HDO molecules given enough time. }
\end{itemize}

In fact, consistent with this conclusion, in the warmer regions surrounding NGC 1333 IRAS2A \citet{Coutens:2014} found a lower \ce{D2O}/\ce{HDO} ratio than in the cold envelope layer around IRAS 16293-2422 \citep{Coutens:2013}. The higher \ce{D2O}/\ce{HDO} than \ce{HDO}/\ce{H2O} ratios reported in both cases are currently under debate. Two possible causes are that either (i) the surface deuteration chemistry network is ill-constrained or (ii) that both sublimation of grain mantles and water formation at higher gas-phase temperatures takes place in the inner regions of this source. Alternatively, if high \ce{D2O} abundances compared to modeled results turn out to be common \citep{Coutens:2013, Coutens:2014}, it must be because the ice mantle does not encounter high temperatures for long enough times. This means that the timescales derived here can also be used in the opposite way, to determine an upper limit for the time that an icy grain resides in an area of a certain temperature. Currently, the limited number of observations does not allow us to draw such a conclusion.

\subsection{Cometary ices}
Shifting focus towards the application of exchange in cometary ices, we note that to the best of our knowledge no large ice chemistry models of comets exist. In the gas-phase coma model of \citet{Rodgers:2002} a high ice abundance is sublimated as a given initial condition. The ice itself, however, is not modeled. They found that the D/H ratio in gas-phase coma species is determined by the ratio in their parents. Thus, the ice in fact does determine the deuterium fractionation found in the coma. Ideally, ice abundances should be determined by a cometary ice chemistry model incorporating low- and high-temperature chemistry, as well as sublimation effects to investigate the effect of accumulative heating on ices.
\begin{itemize}
 \item[] \emph{Thermal chemistry, in particular H/D transfer, can play a major role in the thick cometary ices, changing the deuterium fractionation of many species, because HDO is the main deuterated component of these ices. }
\end{itemize}
{The recent results on the HDO/\ce{H2O} ratio in the coma of comet 67P/Churyumov-Gerasimenko and the discrepancy with respect to the D/H ratios in other Jupiter family comets shows that it is crucial to understand the effect of thermal processing of ice constituents \citep{Altwegg:2014}.}

\subsection{Proof-of-principle modeling}
Scrambling of deuterons at high temperatures can alter the D/H ratios and therefore should be taken into account, when using D/H ratios to determine the cosmic origin of the water on Earth. This holds not only for the main component of interstellar ice, water, but especially also for any species that has an N-H or O-H bond that can participate in hydrogen bonding. Molecules detected in the ISM with such functional groups, most of which are though to have been formed on the surface of dust grains, are \emph{e.g.} \ce{HNO}, \ce{HNC}, \ce{HNCO}, \ce{HNCS}, \ce{NH3}, \ce{NH2OH}, \ce{CH3OH}, \ce{HCONH2}, \ce{CH3NH2}, \ce{NH2CH2CN},  \ce{CH3CH2OH}, \ce{(CH2OH)2}.  
\begin{itemize}
 \item[] \emph{In all examples mentioned in Section~\ref{OtherMol} the hydrogen bonded network for OH and NH moieties plays a crucial role in H/D exchange. Moreover, functional groups with CH that do not participate in hydrogen bonds do not exchange their proton thermally.}
\item[] \emph{We emphasize that as a first approximation exchange of protons and deuterons between water and other hydrogen bonded species should be modeled using an activation energy of $\sim$3840~K both in disk structures as well as in cometary ices.}
\end{itemize}
Note that as discussed above, the forward and backward reaction rates are not expected to be equal, due to a more favorable entropy for the reaction \ce{H2O + D2O -> 2 HDO} with respect to the backward reaction. Although the activation energy itself is constrained only for the case of water and its deuterated analogues, arguments of enthalpy and entropy can also aid in constraining the reaction rates of H/D transfer for other molecules, as suggested by our experiments listed in Table~\ref{exps}. Finally, reactions with no net effect, such as \ce{H2O + HDO -> HDO + H2O}, only have meaning in a microscopic model, where they can be seen as analogues of diffusion. Thus, in a rate equation based model, they can be omitted.

\section{Acknowledgements}
We thank Fabrizio Puletti and Ko-Ju Chuang for their help in the laboratory, Leendertjan Karssemeijer for his help with the hydrogen bonded water structures, and Ewine van Dishoeck, Xander Tielens, Magnus Persson, and Fran\c cois Dulieu for stimulating discussions. 

Astrochemistry in Leiden is supported by the European Community's Seventh Framework Programme (FP7/2007- 2013) under grant agreement n.238258, the Netherlands Research School for Astronomy (NOVA) and from the Netherlands Organization for Scientific Research (NWO) through a VICI grant. T.L. is supported by the Dutch Astrochemistry Network financed by The Netherlands Organization for Scientific Research (NWO). Support for S.I. from the Marie Curie Fellowship (FP7-PEOPLE-2011-IOF-300957) is gratefully acknowledged. H.M.C. is grateful for support from the VIDI research program 700.10.427, which is financed by The Netherlands Organization for Scientific Research (NWO) and from the European Research Council (ERC-2010-StG, Grant Agreement no. 259510-KISMOL).

\bibliographystyle{mn2e}
\bibliography{biblio}

\begin{thebibliography}{}

\bibitem[\protect\citeauthoryear{{Aikawa}, {van Zadelhoff}, {van Dishoeck} \&
  {Herbst}}{{Aikawa} et~al.}{2002}]{Aikawa:2002}
{Aikawa} Y.,  {van Zadelhoff} G.~J.,  {van Dishoeck} E.~F.,    {Herbst} E.,
  2002, \aap, 386, 622

\bibitem[\protect\citeauthoryear{{Albertsson}, {Semenov} \&
  {Henning}}{{Albertsson} et~al.}{2014}]{Albertsson:2014}
{Albertsson} T.,  {Semenov} D.,    {Henning} T.,  2014, \apj, 784, 39

\bibitem[\protect\citeauthoryear{Altwegg, Balsiger, Bar-Nun \& \emph{et.
  al}}{Altwegg et~al.}{2014}]{Altwegg:2014}
Altwegg K.,  Balsiger H.,  Bar-Nun A.,    \emph{et. al} 2014, Science

\bibitem[\protect\citeauthoryear{{Bergin} \& {Tafalla}}{{Bergin} \&
  {Tafalla}}{2007}]{Bergin:2007}
{Bergin} E.~A.,  {Tafalla} M.,  2007, Annu.~ Rev.~ Astron.~ Astrophys., 45, 339

\bibitem[\protect\citeauthoryear{Bertie \& Devlin}{Bertie \&
  Devlin}{1983}]{Bertie:1983}
Bertie J.~E.,  Devlin J.~P.,  1983, The Journal of Chemical Physics, 78

\bibitem[\protect\citeauthoryear{Bertie \& Whalley}{Bertie \&
  Whalley}{1964}]{Bertie:1964}
Bertie J.~E.,  Whalley E.,  1964, The Journal of Chemical Physics, 40

\bibitem[\protect\citeauthoryear{{Bockel{\'e}e-Morvan}, {Lis}, {Wink} \&
  \emph{et al.}}{{Bockel{\'e}e-Morvan} et~al.}{2000}]{Bockelee:2000}
{Bockel{\'e}e-Morvan} D.,  {Lis} D.~C.,  {Wink} J.~E.,    \emph{et al.} 2000,
  \aap, 353, 1101

\bibitem[\protect\citeauthoryear{{Bossa}, {Isokoski}, {de Valois} \&
  {Linnartz}}{{Bossa} et~al.}{2012}]{Bossa:2012}
{Bossa} J.-B.,  {Isokoski} K.,  {de Valois} M.~S.,    {Linnartz} H.,  2012,
  \aap, 545, A82

\bibitem[\protect\citeauthoryear{{Butner}, {Charnley}, {Ceccarelli}, {Rodgers},
  {Pardo}, {Parise}, {Cernicharo} \& {Davis}}{{Butner}
  et~al.}{2007}]{Butner:2007}
{Butner} H.~M.,  {Charnley} S.~B.,  {Ceccarelli} C.,  {Rodgers} S.~D.,  {Pardo}
  J.~R.,  {Parise} B.,  {Cernicharo} J.,    {Davis} G.~R.,  2007, \apjl, 659,
  L137

\bibitem[\protect\citeauthoryear{{Caselli} \& {Ceccarelli}}{{Caselli} \&
  {Ceccarelli}}{2012}]{Caselli:2012}
{Caselli} P.,  {Ceccarelli} C.,  2012, \aapr, 20, 56

\bibitem[\protect\citeauthoryear{Chase}{Chase}{1998}]{Chase:1998}
Chase M.~W.,  1998, Journal of physical and chemical reference data, Monograph,
  no. 9..
American Chemical Society, Washington DC

\bibitem[\protect\citeauthoryear{Cleeves, Bergin, Alexander, Du, Graninger,
  {\o}berg \& Harries}{Cleeves et~al.}{2014}]{Cleeves:2014}
Cleeves L.~I.,  Bergin E.~A.,  Alexander C. M.~O.,  Du F.,  Graninger D.,
  {\o}berg K.~I.,    Harries T.~J.,  2014, Science, 345, 1590

\bibitem[\protect\citeauthoryear{Collier, Ritzhaupt \& Devlin}{Collier
  et~al.}{1984}]{Collier:1984}
Collier W.~B.,  Ritzhaupt G.,    Devlin J.~P.,  1984, The Journal of Physical
  Chemistry, 88, 363

\bibitem[\protect\citeauthoryear{{Congiu}, {Fedoseev}, {Ioppolo}, {Dulieu},
  {Chaabouni}, {Baouche}, {Lemaire}, {Laffon}, {Parent}, {Lamberts}, {Cuppen}
  \& {Linnartz}}{{Congiu} et~al.}{2012}]{Congiu:2012}
{Congiu} E.,  {Fedoseev} G.,  {Ioppolo} S.,  {Dulieu} F.,  {Chaabouni} H.,
  {Baouche} S.,  {Lemaire} J.~L.,  {Laffon} C.,  {Parent} P.,  {Lamberts} T.,
  {Cuppen} H.~M.,    {Linnartz} H.,  2012, \apjl, 750, L12

\bibitem[\protect\citeauthoryear{{Coutens}, {J{\o}rgensen}, {Persson}, {van
  Dishoeck}, {Vastel} \& {Taquet}}{{Coutens} et~al.}{2014}]{Coutens:2014}
{Coutens} A.,  {J{\o}rgensen} J.~K.,  {Persson} M.~V.,  {van Dishoeck} E.~F.,
  {Vastel} C.,    {Taquet} V.,  2014, \apjl, 792, L5

\bibitem[\protect\citeauthoryear{{Coutens}, {Vastel}, {Cazaux}, {Bottinelli},
  {Caux}, {Ceccarelli}, {Demyk}, {Taquet} \& {Wakelam}}{{Coutens}
  et~al.}{2013}]{Coutens:2013}
{Coutens} A.,  {Vastel} C.,  {Cazaux} S.,  {Bottinelli} S.,  {Caux} E.,
  {Ceccarelli} C.,  {Demyk} K.,  {Taquet} V.,    {Wakelam} V.,  2013, \aap,
  553, A75

\bibitem[\protect\citeauthoryear{{Fedoseev}, {Ioppolo}, {Lamberts}, {Zhen},
  {Cuppen} \& {Linnartz}}{{Fedoseev} et~al.}{2012}]{Fedoseev:2012}
{Fedoseev} G.,  {Ioppolo} S.,  {Lamberts} T.,  {Zhen} J.~F.,  {Cuppen} H.~M.,
   {Linnartz} H.,  2012, \jcp, 137, 054714

\bibitem[\protect\citeauthoryear{{Fisher} \& {Devlin}}{{Fisher} \&
  {Devlin}}{1995}]{Fisher:1995}
{Fisher} M.,  {Devlin} P.,  1995, \jpc, 99, 11584

\bibitem[\protect\citeauthoryear{{Fuchs}, {Cuppen}, {Ioppolo}, {Romanzin},
  {Bisschop}, {Andersson}, {van Dishoeck} \& {Linnartz}}{{Fuchs}
  et~al.}{2009}]{Fuchs:2009}
{Fuchs} G.~W.,  {Cuppen} H.~M.,  {Ioppolo} S.,  {Romanzin} C.,  {Bisschop}
  S.~E.,  {Andersson} S.,  {van Dishoeck} E.~F.,    {Linnartz} H.,  2009, \aap,
  505, 629

\bibitem[\protect\citeauthoryear{{Furuya}, {Aikawa}, {Nomura}, {Hersant} \&
  {Wakelam}}{{Furuya} et~al.}{2013}]{Furuya:2013}
{Furuya} K.,  {Aikawa} Y.,  {Nomura} H.,  {Hersant} F.,    {Wakelam} V.,  2013,
  \apj, 779, 11

\bibitem[\protect\citeauthoryear{{G{\'a}lvez}, {Mat{\'e}}, {Herrero} \&
  {Escribano}}{{G{\'a}lvez} et~al.}{2011}]{Galvez:2011}
{G{\'a}lvez} {\'O}.,  {Mat{\'e}} B.,  {Herrero} V.~J.,    {Escribano} R.,
  2011, \apj, 738, 133

\bibitem[\protect\citeauthoryear{{Grim}, {Greenberg}, {de Groot}, {Baas},
  {Schutte} \& {Schmitt}}{{Grim} et~al.}{1989}]{Grim:1989}
{Grim} R.~J.~A.,  {Greenberg} J.~M.,  {de Groot} M.~S.,  {Baas} F.,  {Schutte}
  W.~A.,    {Schmitt} B.,  1989, Astron.~Astrophys.~Suppl.~Ser., 78, 161

\bibitem[\protect\citeauthoryear{{Hasegawa}, {Herbst} \& {Leung}}{{Hasegawa}
  et~al.}{1992}]{Hasegawa:1992}
{Hasegawa} T.~I.,  {Herbst} E.,    {Leung} C.~M.,  1992,
  Astrophys.~J.~Suppl.~Ser., 82, 167

\bibitem[\protect\citeauthoryear{Hiraoka, Miyagoshi, Takayama, Yamamoto \&
  Kihara}{Hiraoka et~al.}{1998}]{Hiraoka:1998}
Hiraoka K.,  Miyagoshi T.,  Takayama T.,  Yamamoto K.,    Kihara Y.,  1998,
  Astrophys.~J.~, 498, 710

\bibitem[\protect\citeauthoryear{Ioppolo, Fedoseev, Lamberts, Romanzin \&
  Linnartz}{Ioppolo et~al.}{2013}]{Ioppolo:2013}
Ioppolo I.,  Fedoseev G.,  Lamberts T.,  Romanzin C.,    Linnartz H.,  2013,
  Rev.~Sci.~Instrum.~, 84, 073112

\bibitem[\protect\citeauthoryear{{Ioppolo}, {Cuppen}, {Romanzin}, {van
  Dishoeck} \& {Linnartz}}{{Ioppolo} et~al.}{2008}]{Ioppolo:2008}
{Ioppolo} S.,  {Cuppen} H.~M.,  {Romanzin} C.,  {van Dishoeck} E.~F.,
  {Linnartz} H.,  2008, Astrophys.~ J., 686, 1474

\bibitem[\protect\citeauthoryear{Itikawa \& Mason}{Itikawa \&
  Mason}{2005}]{Itikawa:2005}
Itikawa Y.,  Mason N.,  2005, Journal of Physical and Chemical Reference Data,
  34

\bibitem[\protect\citeauthoryear{{J{\o}rgensen}, {Sch{\"o}ier} \& {van
  Dishoeck}}{{J{\o}rgensen} et~al.}{2002}]{Jorgensen:2002}
{J{\o}rgensen} J.~K.,  {Sch{\"o}ier} F.~L.,    {van Dishoeck} E.~F.,  2002,
  \aap, 389, 908

\bibitem[\protect\citeauthoryear{Jung, Park, Kim \& Kang}{Jung
  et~al.}{2004}]{Jung:2004}
Jung K.-H.,  Park S.-C.,  Kim J.-H.,    Kang H.,  2004, \jcp, 121, 2758

\bibitem[\protect\citeauthoryear{Karssemeijer}{Karssemeijer}{2014}]{Karssemeij%
er:2014pc}
Karssemeijer L.~J.,  2014, Personal communication, Based on the ASW samples
  used in \citep{Karssemeijer:2014}

\bibitem[\protect\citeauthoryear{{Karssemeijer}, {Ioppolo}, {van Hemert}, {van
  der Avoird}, {Allodi}, {Blake} \& {Cuppen}}{{Karssemeijer}
  et~al.}{2014}]{Karssemeijer:2014}
{Karssemeijer} L.~J.,  {Ioppolo} S.,  {van Hemert} M.~C.,  {van der Avoird} A.,
   {Allodi} M.~A.,  {Blake} G.~A.,    {Cuppen} H.~M.,  2014, \apj, 781, 16

\bibitem[\protect\citeauthoryear{Kim, Kim \& Kang}{Kim et~al.}{2009}]{Kim:2009}
Kim J.-H.,  Kim Y.-K.,    Kang H.,  2009, The Journal of Chemical Physics, 131,

\bibitem[\protect\citeauthoryear{{Launhardt}, {Stutz}, {Schmiedeke}, {Henning},
  {Krause} \& {\emph{et al.}}}{{Launhardt} et~al.}{2013}]{Launhardt:2013}
{Launhardt} R.,  {Stutz} A.~M.,  {Schmiedeke} A.,  {Henning} T.,  {Krause} O.,
    {\emph{et al.}} 2013, \aap, 551, A98

\bibitem[\protect\citeauthoryear{Lee, Lee, Kim \& Kang}{Lee
  et~al.}{2007}]{Lee:2007}
Lee C.-W.,  Lee P.-R.,  Kim Y.-K.,    Kang H.,  2007, The Journal of Chemical
  Physics, 127,

\bibitem[\protect\citeauthoryear{Miyauchi, Hidaka, Chigai, Nagaoka, Watanabe \&
  Kouchi}{Miyauchi et~al.}{2008}]{Miyauchi:2008}
Miyauchi N.,  Hidaka H.,  Chigai T.,  Nagaoka A.,  Watanabe N.,    Kouchi A.,
  2008, Chem.~Phys.~Lett., 456, 27

\bibitem[\protect\citeauthoryear{Moon, Yoon \& Kang}{Moon
  et~al.}{2010}]{Moon:2010}
Moon E.-S.,  Yoon J.,    Kang H.,  2010, \jcp, 133, 044709

\bibitem[\protect\citeauthoryear{{Morbidelli}, {Chambers}, {Lunine}, {Petit},
  {Robert}, {Valsecchi} \& {Cyr}}{{Morbidelli} et~al.}{2000}]{Morbidelli:2000}
{Morbidelli} A.,  {Chambers} J.,  {Lunine} J.~I.,  {Petit} J.~M.,  {Robert} F.,
   {Valsecchi} G.~B.,    {Cyr} K.~E.,  2000, Meteoritics and Planetary Science,
  35, 1309

\bibitem[\protect\citeauthoryear{Muralidharan, Deymier, Stimpfl, de Leeuw \&
  Drake}{Muralidharan et~al.}{2008}]{Muralidharan:2008}
Muralidharan K.,  Deymier P.,  Stimpfl M.,  de Leeuw N.~H.,    Drake M.~J.,
  2008, Icarus, 198, 400

\bibitem[\protect\citeauthoryear{{Nagaoka}, {Watanabe} \& {Kouchi}}{{Nagaoka}
  et~al.}{2005}]{Nagaoka:2005}
{Nagaoka} A.,  {Watanabe} N.,    {Kouchi} A.,  2005, \apjl, 624, L29

\bibitem[\protect\citeauthoryear{{Nomura} \& {Millar}}{{Nomura} \&
  {Millar}}{2005}]{Nomura:2005}
{Nomura} H.,  {Millar} T.~J.,  2005, \aap, 438, 923

\bibitem[\protect\citeauthoryear{Oba, Miyauchi, Hidaka, Chigai, Watanabe \&
  Kouchi}{Oba et~al.}{2009}]{Oba:2009}
Oba Y.,  Miyauchi N.,  Hidaka H.,  Chigai T.,  Watanabe N.,    Kouchi A.,
  2009, Astrophys.~J., 701, 464

\bibitem[\protect\citeauthoryear{O'Brien, Walsh, Morbidelli, Raymond \&
  Mandell}{O'Brien et~al.}{2014}]{OBrien:2014}
O'Brien D.~P.,  Walsh K.~J.,  Morbidelli A.,  Raymond S.~N.,    Mandell A.~M.,
  2014, Icarus, 239, 74

\bibitem[\protect\citeauthoryear{Oxley, Zahn \& Pursell}{Oxley
  et~al.}{2006}]{Oxley:2006}
Oxley S.~P.,  Zahn C.~M.,    Pursell C.~J.,  2006, The Journal of Physical
  Chemistry A, 110, 11064

\bibitem[\protect\citeauthoryear{{Palumbo}}{{Palumbo}}{2006}]{Palumbo:2006}
{Palumbo} M.~E.,  2006, \aap, 453, 903

\bibitem[\protect\citeauthoryear{Park, Jung \& Kang}{Park
  et~al.}{2004}]{Park:2004}
Park S.-C.,  Jung K.-H.,    Kang H.,  2004, The Journal of Chemical Physics,
  121

\bibitem[\protect\citeauthoryear{Ratajczak}{Ratajczak}{2012}]{Ratajczak:2012}
Ratajczak A.,  2012, Theses, {Universit{\'e} de Grenoble}

\bibitem[\protect\citeauthoryear{{Ratajczak}, {Quirico}, {Faure}, {Schmitt} \&
  {Ceccarelli}}{{Ratajczak} et~al.}{2009}]{Ratajczak:2009}
{Ratajczak} A.,  {Quirico} E.,  {Faure} A.,  {Schmitt} B.,    {Ceccarelli} C.,
  2009, \aap, 496, L21

\bibitem[\protect\citeauthoryear{{Ratajczak}, {Taquet}, {Kahane}, {Ceccarelli},
  {Faure} \& {Quirico}}{{Ratajczak} et~al.}{2011}]{Ratajczak:2011}
{Ratajczak} A.,  {Taquet} V.,  {Kahane} C.,  {Ceccarelli} C.,  {Faure} A.,
  {Quirico} E.,  2011, \aap, 528, L13

\bibitem[\protect\citeauthoryear{{Rodgers} \& {Charnley}}{{Rodgers} \&
  {Charnley}}{2002}]{Rodgers:2002}
{Rodgers} S.~D.,  {Charnley} S.~B.,  2002, \mnras, 330, 660

\bibitem[\protect\citeauthoryear{{Sch{\"o}ier}, {J{\o}rgensen}, {van Dishoeck}
  \& {Blake}}{{Sch{\"o}ier} et~al.}{2002}]{Schoier:2002}
{Sch{\"o}ier} F.~L.,  {J{\o}rgensen} J.~K.,  {van Dishoeck} E.~F.,    {Blake}
  G.~A.,  2002, \aap, 390, 1001

\bibitem[\protect\citeauthoryear{Souda, Kawanowa, Kondo \& Gotoh}{Souda
  et~al.}{2003}]{Souda:2003}
Souda R.,  Kawanowa H.,  Kondo M.,    Gotoh Y.,  2003, The Journal of Chemical
  Physics, 119

\bibitem[\protect\citeauthoryear{Straub, Lindsay, Smith \& Stebbings}{Straub
  et~al.}{1998}]{Straub:1998}
Straub H.~C.,  Lindsay B.~G.,  Smith K.~A.,    Stebbings R.~F.,  1998, The
  Journal of Chemical Physics, 108

\bibitem[\protect\citeauthoryear{Thornton, Khatkale \& Devlin}{Thornton
  et~al.}{1981}]{Thornton:1981}
Thornton C.,  Khatkale M.~S.,    Devlin J.~P.,  1981, The Journal of Chemical
  Physics, 75

\bibitem[\protect\citeauthoryear{{Tielens}}{{Tielens}}{1983}]{Tielens:1983}
{Tielens} A.~G.~G.~M.,  1983, \aap, 119, 177

\bibitem[\protect\citeauthoryear{Tielens}{Tielens}{2013}]{Tielens:2013}
Tielens A. G. G.~M.,  2013, Rev. Mod. Phys., 85, 1021

\bibitem[\protect\citeauthoryear{Uras-Aytemiz, Joyce \& Devlin}{Uras-Aytemiz
  et~al.}{2001}]{Uras:2001}
Uras-Aytemiz N.,  Joyce C.,    Devlin J.~P.,  2001, The Journal of Chemical
  Physics, 115

\bibitem[\protect\citeauthoryear{{van der Tak}, {van Dishoeck}, {Evans} II \&
  {Blake}}{{van der Tak} et~al.}{2000}]{Tak:2000}
{van der Tak} F.~F.~S.,  {van Dishoeck} E.~F.,  {Evans} II N.~J.,    {Blake}
  G.~A.,  2000, \apj, 537, 283

\bibitem[\protect\citeauthoryear{{Vastel}, {Ceccarelli}, {Caux}, {Coutens} \&
  {\emph{et al.}}}{{Vastel} et~al.}{2010}]{Vastel:2010}
{Vastel} C.,  {Ceccarelli} C.,  {Caux} E.,  {Coutens} A.,    {\emph{et al.}}
  2010, \aap, 521, L31

\bibitem[\protect\citeauthoryear{{Walsh}, {Millar} \& {Nomura}}{{Walsh}
  et~al.}{2010}]{Walsh:2010}
{Walsh} C.,  {Millar} T.~J.,    {Nomura} H.,  2010, \apj, 722, 1607

\bibitem[\protect\citeauthoryear{{Watanabe} \& {Kouchi}}{{Watanabe} \&
  {Kouchi}}{2002}]{Watanabe:2002}
{Watanabe} N.,  {Kouchi} A.,  2002, \apjl, 571, L173

\bibitem[\protect\citeauthoryear{Wooldridge \& Devlin}{Wooldridge \&
  Devlin}{1988}]{Woolridge:1988}
Wooldridge P.~J.,  Devlin J.~P.,  1988, The Journal of Chemical Physics, 88

\end{thebibliography}

\end{document}